\documentclass[reprint, superscriptaddress, aps, showkeys, pra]{revtex4-2}

\usepackage[colorlinks=true, citecolor=blue, urlcolor=blue, linkcolor=magenta]{hyperref}
\usepackage{braket}
\usepackage{amsmath,amssymb,amsthm}
\usepackage{easybmat}
\usepackage[colorlinks=true,citecolor=blue,urlcolor=blue, linkcolor=magenta]{hyperref}
\usepackage[pdftex]{graphicx}
\usepackage{times,txfonts}
\usepackage{braket}
\usepackage{color}
\usepackage{natbib}
\usepackage{appendix}

\begin{document}
    \title{Quantum speed limit and nonclassicality in open quantum system models using the Wigner function}
\author{Arti Gaharwar\textsuperscript{}}
\email{artigaharwar952@gmail.com}
\author{Devvrat Tiwari\textsuperscript{}}
\email{devvrat.1@iitj.ac.in}
\author{Subhashish Banerjee\textsuperscript{}}
\email{subhashish@iitj.ac.in}
\affiliation{Indian Institute of Technology Jodhpur-342030, India\textsuperscript{}}

\date{\today}

\begin{abstract}
    The quantum speed limit and the Wigner function of open system models are studied. To this end, we use the phase covariant and a two-qubit model interacting with a squeezed thermal bath via position-dependent coupling. The dependence of the coupling on the position of the qubits allows for the study of the dynamics in the collective regime, which is conducive to speeding up the evolution. An interesting interplay is observed between non-Markovian behavior, quantumness, and the quantum speed limit. The presence of quantum correlations is seen to speed up the evolution.   
\end{abstract}

\maketitle
\section{Introduction}

Open quantum systems and non-Markovian dynamics ~\cite{banerjee2018open,breuer2002theory} are fundamental concepts that underpin quantum mechanics, offering insights into the behavior of quantum systems in interaction with their environment. Unlike closed quantum systems, which evolve unitarily according to Schrödinger's equation, open quantum systems undergo non-unitary dynamics characterized by decoherence and dissipation due to external interactions ~\cite{breuer2016colloquium,zhang2012general}. These systems are pivotal across various domains, including quantum information processing, quantum optics, condensed matter physics, and quantum thermodynamics~\cite{carmichael2009open,daley2014quantum,sieberer2016keldysh,kosloff2013quantum}. Non-Markovian evolution~\cite{nonmarkovian1,nonmarkovian2}, in contrast to the Markovian paradigm, introduces memory effects where a system's past trajectory influences its current state, arising from strong environmental correlations, intricate coupling mechanisms, or long-range interactions. Understanding and characterizing open quantum systems with non-Markovian dynamics is critical for unraveling complex quantum phenomena, designing robust quantum technologies, and exploring the interplay between quantum systems and their environments~\cite{acin2018quantum,kurizki2015quantum,koch2016controlling}. 

The quantum speed limit (QSL) bounds the evolution of quantum systems. This bound was originally derived from the energy-time uncertainty principle~\cite{uncertainty}. Initially, a bound on the speed of evolution was derived for dynamics between orthogonal states for isolated systems. 
The speed limit for driven quantum systems that are valid for arbitrary initial and final states was derived in~\cite{zhang2012general,hegerfeldt2013driving}. A related concept is the QSL time, whose unified lower bound for closed systems with unitary evolution was obtained by Mandelstam-Tamm (MT)~\cite{mandelstam1945uncertainty} and Margolus-Levitin (ML)~\cite{margolus1998maximum}. The extensions of the MT and ML bounds to the non-orthogonal states and to driven systems have been investigated in~\cite{sun2019distinct,ness2022quantum,haseli2020controlling}. QSL time using the Fubini-Study metric on the space of pure quantum states was given in~\cite{anandan_aharanov}. This led to the utilization of geometric measures in an extensive study of QSL time in recent decades~\cite{PATI1991105, ANANDAN199729, Giovannetti_2003, Deffner_2017}.
Later, the concept of the QSL was generalized for open quantum systems~\cite{QSLinOQS1, QSLinOQS2}. Further, the notion of QSL time was explored in the scenarios of non-uniform magnetic field~\cite{Aggarwal_2022}, phase covariant dynamics~\cite{Phasecovariant}, spin baths~\cite{Devvrat_QSL}, quantum thermodynamics and quantum batteries~\cite{Campo2014, Lewenstein_2020, Pati_2020, Devvrat2023}, as well as in neutrinos~\cite{bouri2024probing} and neutral meson oscillation dynamics~\cite{Banerjee2023}, among others. It has found a wide range of applications in tasks related to quantum information theory~\cite{Nori_2012, Bekenstein_1981, Paulson2022}, quantum computation~\cite{Lloyd2000}, metrology~\cite{Kok_2010, Giovannetti2011}, quantum optimal control algorithms~\cite{Santoro_2009}, and quantum gravity~\cite{Liegener_2022, Heng_2023}.

The dynamics of classical systems are easily accessible using phase space techniques, which is denied in the quantum case because of the uncertainty principle. However, it is still possible to construct quasi-probability distributions (QDs)~\cite{STRATONOVICH, Agarwal_2012, Quasiprobability, THAPLIYAL2016148} for quantum mechanical systems analogous to their classical counterparts. Wigner derived the first QD, now known as the Wigner function~\cite{wigner1, Wigner2}, where phase space representation was brought into the field of quantum mechanics. However, this differs from the usual probability distributions, as it may take negative values. This negative value, in general, is used as a witness of quantumness in a system~\cite{arkhipov2018negativity,semenov2006nonclassicality,genoni2013detecting}. 

The choice of a measure for the distinguishability of quantum states determines the actual mathematical representation of the fundamental limit of the speed of evolution. Many different measures have been developed over the years, such as the Fisher information measure~\cite{Deffner_2017}, quantum relative purity~\cite{QSLinOQS1}, and Bures angle metric~\cite{QSLinOQS2}. The Schatten-p-norm of the generator of quantum dynamics qualitatively governs quantum speed limits. Since computing Schatten-p-norms can be mathematically complex, a different method utilizing the Wigner function was proposed in~\cite{Geometricquantumspeed, PhysRevLett.120.070401}. It was found that the QSL in Wigner space is equivalent to expressions in density operator space but that the new bound is easier to compute. Further, it facilitates discussion on nonclassicality and quantum speed limit on a similar footing. Recently, the QSL was also computed using the Kirkwood-Dirac quasiprobabilities~\cite{pratapsi2024}, in Liouville space for open quantum systems~\cite{Uzdin_2016}, and via the trajectory ensemble~\cite{Xu_2020}.

In this work, we explore the QSL and the Wigner function of the single and two-qubit open system models. Together with this, we study the regions of nonclassicality in a quantum system using the Wigner function and the corresponding nonclassical volume. The single qubit system is evolved using the phase covariant channel~\cite{UltimatePrecisionLimits, phase123, Filippov2020}. This allows for a consideration of unital and non-unital effects on the same platform along with (non-)Markovian evolution. Further, the two-qubit model, interacting with a squeezed thermal bath, is characterized by the inter-qubit distance~\cite{thermalbath, collectivenonclassical, Banerjee2009}. This allows the dynamics to be considered in two regimes, that is, the independent and the collective decoherence regimes. In the collective regime, the inter-qubit distance is smaller compared to the environment's length scale, allowing for interesting features. We explore the connection of the QSL for this model with the quantum correlations. 

The article is arranged as follows. In Sec.~\ref{prelim}, we discuss the preliminaries used throughout the paper, namely, the Wigner function and non-classical volume, the QSL, the phase covariant channel, and the two-qubit collective decoherence model. The dynamics of the Wigner function, non-classical volume, and the QSL for the above two models are studied in Sec.~\ref{analysis}, followed by the conclusions in Sec.~\ref{conclusion}.  

\section{Preliminaries}\label{prelim}

Here, we briefly discuss the Wigner function and the quantum speed limit (QSL) in the Wigner phase space. Further, we discuss the phase covariant channel and a two-qubit dissipative model. 
\subsection{Wigner function}

For a single spin-$j$ state, the Wigner function~\cite{wigner_func, Quasiprobability} is given by
\begin{equation}\label{eq_wigner_function}
W(\theta, \phi) = \sqrt{\frac{2j + 1}{4\pi}} \sum_{K,Q} \rho_{KQ} Y_{KQ}(\theta, \phi),
\end{equation}
where $K = 0,1,.... 2j$, $Q = -K , -K+1,...,0,...,K-1,K$, and 
\begin{equation}
\rho_{KQ} = \text{Tr} \left( {T}^\dagger_{KQ} \rho \right).
\end{equation}
Further, ${Y}_{KQ}$ are the spherical harmonics, and ${T}_{KQ}$ are multipole operators given by
\[
{T}_{KQ} = \sum_{m,m'} (-1)^{j-m} \sqrt{2K + 1} \begin{pmatrix} j & K & j \\ -m & Q & m' \end{pmatrix} |j, m\rangle \langle j, m'|,
\]
where $\begin{pmatrix}j_1 &j_2& j \\ m_1& m_2& m\end{pmatrix} = \frac{(-1)^{j_1 - j_2 - m}} {\sqrt{2j + 1}} \langle j_1 m_1 j_2 m_2 | j - m \rangle$ is the Wigner-$3j$ symbol~\cite{angular}, and $\langle j_1 m_1 j_2 m_2 | j - m \rangle$ is the Clebsch-Gordan coefficient. The multipole operators for $j=\frac{1}{2}$ are discussed in~\cite{Quasiprobability}. The Wigner function follows the normalization condition
\begin{equation}
\int\limits_{0}^{2\pi} \int\limits_{0}^{\pi} W(\theta, \phi) \sin \theta d\theta d\phi = 1,
\end{equation}
and $W^*(\theta, \phi) = W(\theta, \phi)$. Moreover, the Wigner function of a system composed of two spin-$j$ particles is given by
\begin{align}
W(\theta_1, \phi_1, \theta_2, \phi_2) &=  \frac{2j + 1}{4\pi} \sum_{K_1,Q_1} \sum_{K_2,Q_2} \rho_{K_1Q_1K_2Q_2} Y_{K_1Q_1}(\theta_1, \phi_1) \nonumber \\
&\times Y_{K_2Q_2}(\theta_2, \phi_2),
\end{align}
where $\rho_{K_1Q_1K_2Q_2} = \text{Tr} \left( \rho {T}^\dagger_{K_1Q_1} {T}^\dagger_{K_2Q_2} \right)$, satisfying the normalization condition
\begin{align}
\int\limits_{0}^{2\pi} \int\limits_{0}^{\pi} \int\limits_{0}^{2\pi} \int\limits_{0}^{\pi} W(\theta_1, \phi_1, \theta_2, \phi_2) \sin\theta_1 \sin\theta_2 \,d\theta_1 \,d\phi_1 \,d\theta_2 \,d\phi_2 = 1.
\end{align}
\subsection{Nonclassical volume}
Negative values of the Wigner function provide a signature of nonclassicality. However, the negative values do not provide a quantitative measure of nonclassicality. A measure of nonclassicality is the nonclassical volume, introduced in~\cite{Anatole_Kenfack_2004}. The definition of the nonclassical volume is given by 
\begin{align}
\delta = \int\int |W(\theta, \phi)| \sin\theta \, d\theta \, d\phi - 1. 
\label{nonclassical volume}\end{align}
It can be observed that a non-zero value of $\delta$ implies the existence of nonclassicality in the system. 

\subsection{Wasserstein-1-distance and QSL}
The Wasserstein-1-distance measures the total variation distance between two probability distributions. It quantifies the highest absolute difference in probabilities assigned to the same event by the two distributions~\cite{wasserstein_cite}. The expression for the Wasserstein-1-distance, which can be regarded as a generalization of the trace distance to quasiprobability distributions, is given by
\begin{align}\label{Wasserstein_1_distance_def}
D(W(t), W(0)) &= \left\| W(t) - W(0) \right\|_1\nonumber \\
&= \int |W(\theta, \phi,t) - W(\theta, \phi, 0)| \, \sin(\theta)d\theta d\phi,
\end{align}
where ($\theta,\phi$) are the phase space coordinates. Using the Wasserstein-1 distance, an expression for the QSL $v_{\text{QSL}}^W$ in the phase space was developed in~\cite{Geometricquantumspeed}, which is given by
\begin{equation}
v_{QSL}^W \equiv \min \left\{ \left\| \dot{W}(t) \right\|_p \text{ for } p \in [0, \infty) \right\},
\label{eq_vqsl}
\end{equation}
where $\left\|\dot{W}(t)\right\|_p = \left(\int \left|\dot{W}(\theta, \phi, t)\right|^p \sin(\theta) d\theta d\phi\right)^{1/p}$.
From above, the QSL time in the Wigner phase space can be shown to be
\begin{equation}\label{eq_tau_qsl}
\tau_{QSL}^W \equiv \frac{D\left(W(\tau), W(0)\right)}{\frac{1}{\tau} \int_{0}^{\tau} dt \,v_{QSL}^W},
\end{equation}
where $\tau$ is the actual driving time. 

\subsection{Phase covariant channel}\label{phase_covariant_channel}
The dynamics of a single qubit system under the action of a phase covariant channel is given by the master equation of the form~\cite{UltimatePrecisionLimits, Filippov2020,phase123}
\begin{align}
\frac{d\rho}{dt} &= \frac{\gamma_1(t)}{2} \mathcal{L}_1[\rho(t)] + \frac{\gamma_2(t)}{2} \mathcal{L}_2[\rho(t)] + \frac{\gamma_3(t)}{2} \mathcal{L}_3[\rho(t)],
\label{phase_master_equation}
\end{align}
where
\begin{align}
\mathcal{L}_1(\rho(t)) &= \sigma_+\rho(t)\sigma_- - \frac{1}{2}\left\{\sigma_-\sigma_+, \rho(t)\right\}, \nonumber \\
\mathcal{L}_2(\rho(t)) &= \sigma_-\rho(t)\sigma_+ - \frac{1}{2}\left\{\sigma_+\sigma_-, \rho(t)\right\}, \nonumber \\
\mathcal{L}_3(\rho(t)) &= \sigma_z\rho(t)\sigma_z - \rho(t),
\end{align}
and
\begin{equation}
\sigma_\pm = \frac{\sigma_x \pm i\sigma_y}{2}.
\end{equation}
Further, \(\sigma_x\) and \(\sigma_y\) are the Pauli spin matrices. Let $\Phi_t$ be the phase covariant map acting on a single qubit density matrix $\rho(0) = \frac{1}{2}\begin{pmatrix}
    1 + z(0) & x(0) - iy(0)\\ x(0) + iy(0) & 1 - z(0)
\end{pmatrix}$, where $k(0) = {\rm Tr}[\sigma_k\rho(0)]$ (for $k = x, y, z$). The action of this map leads to a single qubit density matrix $\rho(t)$ given by~\cite{Filippov2020, phase123}
\begin{align}
\rho(t) &= \Phi_t\left[\rho(0)\right] \nonumber \\
&= 
\frac{1}{2}\begin{pmatrix}
    1 + G(t) + z(0)\lambda_z(t) & \lambda_x(t) [x(0) - iy(0)] \\
    \lambda_x(t) [x(0) + iy(0)] & 1 - G(t) - z(0)\lambda_z(t) \\
\end{pmatrix},\label{eq:density}
\end{align}
where
\begin{align}
G(t) &= e^{-\Gamma_1(t) -\Gamma_2(t)}\int_0^t [\gamma_1(s) - \gamma_2(s)] e^{\Gamma_1(s) + \Gamma_2(s)}ds, \nonumber \\
\lambda_x(t) &= e^{-\frac{\Gamma_1(t)}{2} -\frac{\Gamma_2(t)}{2} -2\Gamma_3(t) }, ~~\text{and}\nonumber \\
\lambda_z(t) &= e^{-\Gamma_1(t) - \Gamma_2(t)},
\end{align}
such that $\Gamma_i = \int_0^t \gamma_i(s)ds$ for $i = 1,2,3$.
The framework of a phase covariant map, Eq.~(\ref{phase_master_equation}), addresses the evolution of quantum states under different physical processes, such as absorption, emission, and dephasing, which are represented by rate constants $\gamma_1(t)$, $\gamma_2(t)$, and $\gamma_3(t)$, respectively. In our investigation, we consider a combination of the non-Markovian amplitude damping (NMAD) channel and the non-Markovian random telegraph noise (NMRTN) pure dephasing channel. The NMAD channel models physical processes such as spontaneous emission, while in the NMRTN channel, the decoherence processes result from low-frequency noise~\cite{nonmarkovian2, Phasecovariant}. In the present context, we set the absorption coefficient to be zero ($\gamma_1(t)=0$), and both emission and dephasing coefficients are non-zero.
For the NMAD channel, we have
\begin{equation}
\gamma_2(t) = \frac{4\kappa l \sinh\left(\frac{zt}{2}\right)}{z \cosh\left(\frac{zt}{2}\right) + l \sinh\left(\frac{zt}{2}\right)},
\end{equation}
where  $z = \sqrt{l^2 - 2\kappa l}$. $\kappa$ is the qubit-environment coupling strength, and $l$ is the spectral width related to the reservoir correlation time. In the case of the NMAD channel, \(l > 2\kappa\) is the region of Markovian dynamics, whereas the region for non-Markovian dynamics is \(l < 2\kappa\). Furthermore, for the NMRTN channel, the dephasing rate is given by 
\begin{equation}
\gamma_3(t) = \frac{\eta(\mu^2 + 1) \sin(\mu \eta t)}{\mu \cos(\mu \eta t) + \sin(\mu \eta t)},
\end{equation}
where \( \mu = \sqrt{\left(\frac{2\nu}{\eta}\right)^2 - 1} \). Here, \( \eta \) is the spectral bandwidth, and \( \nu \) is the coupling strength between the qubit and the reservoir. In the case of the NMRTN channel, we have non-Markovian dynamics for \( \left(\frac{2\nu}{\eta}\right)^2> 1 \) and Markovian dynamics for  \(\left(\frac{2\nu}{\eta}\right)^2 < 1 \).

\subsection{Two-qubit dissipative interaction with squeezed thermal bath}
The Hamiltonian for dissipative interaction of $N$ two-level atoms (qubits) with the bath (modeled as a 3-D electromagnetic field) via the dipole interaction was discussed in~\cite{thermalbath,collectivenonclassical}. This is given by 
\begin{align}
    H &= H_S + H_R + H_{SR} \\
    H &= \sum_{n=1}^{N} \hbar\omega_{n} S_n^z + \sum_{\Vec{k}s} \sum_{n=1}^{N} \hbar\omega_{k}\left(b_{\Vec{k}s}^\dagger b_{\Vec{k}s}+\frac{1}{2}\right) \nonumber \\
    &- i\hbar \sum_{\Vec{k}s} \sum_{n=1}^{N}\left[ \Vec{\mu_{n}} \cdot \Vec{g}_{\vec{k}s}\left(\Vec{r_{n}}\right)\left( S_n^+ +S_n^-\right)b_{\Vec{k}s}-h.c\right],
    \label{total_Hamiltonian_2_qubit}
\end{align}
where
$S_n^z =\frac{1}{2}\left(\ket{e^n}\bra{e^n}-\ket{g^n}\bra{g^n}\right)$
is the energy operator of $n^{th}$ atom while  $b_{\Vec{k}s}^\dagger$ and $b_{\Vec{k}s}$ are the creation and annihilation operators of the field mode and 
$\Vec{g}_{\vec{k}s}\left(\Vec{r_{n}}\right)=\left(\frac{\omega_n}{2\hbar \epsilon V}\right)\vec{e}_{\Vec{k}s}e^{i\Vec{k}.r_n}$ is system–reservoir coupling constant where V is normalization volume and $\vec{e}_{\Vec{k}s}$ is the unit polarization vector of the field. A similar discussion for non-dissipative interaction was made in~\cite{Banerjee2009}.

Here, we consider a two-qubit system ($N=2$) interacting with a 3-D electromagnetic field initially in the squeezed thermal state.
The master equation for the reduced density matrix of this system, under the usual Born–Markov and rotating wave approximation (RWA), is given by
\begin{align}
\frac{d\rho}{dt} = &-\frac{i}{\hbar} \left[ \tilde H_S, ~\rho \right] \nonumber \\
&- \frac{1}{2} \sum_{i,j=1}^{2} \Gamma_{ij} \left( {1 + N_e} \right) \left( \rho S_+^i S_-^j + S_+^i S_-^j\rho-2 S_-^j\rho  S_+^i \right) \nonumber \\
&- \frac{1}{2} \sum_{i,j=1}^{2} \Gamma_{ij} N_e \left( \rho S_-^i S_+^j + S_-^i S_+^j\rho-2 S_+^j\rho  S_-^i \right) \nonumber \\
&+ \frac{1}{2} \sum_{i,j=1}^{2} \Gamma_{ij} M_e \left( \rho S_+^i S_+^j + S_+^i S_+^j\rho -2 S_+^j\rho  S_+^i\right) \nonumber \\
&+\frac{1}{2} \sum_{i,j=1}^{2} \Gamma_{ij} M_{e}^* \left( \rho S_-^i S_-^j + S_-^i S_-^j\rho -2 S_-^j\rho  S_-^i\right), \label{eq:master_equation}
\end{align}
where $ S_-^n= \ket{g^n}\bra{e^n}$ and  $S_+^n= | e^n \rangle\langle g^n|$ are dipole lowering and raising operators, respectively. Further, 
\begin{align}
N_e &= N_{\text{th}} \left( \cosh^2(r) + \sinh^2(r) \right) + \sinh^2(r), \\
M_e &= -\frac{1}{2} \sinh(2r) e^{i\phi}(2N_{\text{th}} + 1) \equiv R e^{i\phi(\omega_0)},
\end{align}
with \( \omega_0 = \frac{\omega_1 + \omega_2}{2} \). \( N_{\text{th}} = \frac{1}{e^{\frac{\hbar \omega}{k_B T}} - 1}\) is the Planck distribution giving the number of thermal photons at the frequency \( \omega \), and temperature \( T \) and \( r, \phi \) are the bath squeezing parameters. By putting these squeezing parameters to zero ($r = \phi = 0$), one can obtain a thermal bath without squeezing, and further, by setting $T=0$, one gets the vacuum bath scenario. Here, we focus on the dynamics of an initial two-qubit state under the influence of a squeezed thermal bath. In the above master equation,
\begin{align}
\tilde H_S &= \hbar \sum_{n=1}^{2} \omega_n S_z^n + \hbar \sum_{i,j=1}^{2} \Omega_{ij} S_+^i S_-^j, ~~\text{where}~~ (i \neq j),\\
\Omega_{ij} &= \frac{3}{4} \sqrt{\frac{\Gamma_i \Gamma_j}{2}} \left( \left[ (\mu \cdot r_{ij})^2 - 1 \right] \frac{\cos(k_0 r_{ij})}{k_0 r_{ij}} \right. \nonumber \\
& + \left. \left[1 - 3(\mu \cdot r_{ij})^2\right] \left[\frac{\sin(k_0 r_{ij})}{(k_0 r_{ij})^2} + \frac{\cos(k_0 r_{ij})}{(k_0 r_{ij})^3}\right] \right), ~~\text{and}\\
F(k_0 r_{ij}) &= \frac{3}{2} \left( \left[1 - (\mu \cdot r_{ij})^2 \right] \frac{\sin(k_0 r_{ij})}{k_0 r_{ij}} \right. \nonumber \\
&\quad + \left. \left[1 - 3(\mu \cdot r_{ij})^2\right] \left[\frac{\cos(k_0 r_{ij})}{(k_0 r_{ij})^2} + \frac{\sin(k_0 r_{ij})}{(k_0 r_{ij})^3}\right] \right). 
\end{align}%
Here, $S_z^n$ is the energy operator of the $n$th atom, and $\mu=\mu_1=\mu_2$ are the transition dipole moments, dependent on the different atomic positions and $k_0=\frac{\omega_0}{c}$. The wavevector $k_0=\frac{2\pi}{\lambda_0}$, where $\lambda_0$ is the resonant wavelength. Now, $k_0. r_{ij}$ is the ratio between the interatomic distance and the resonant wavelength. This ratio enables us to talk about the dynamics in the regime of independent decoherence where $\frac{ r_{ij}}{\lambda_0} \geq1$ and in the regime of collective decoherence where  $\frac{ r_{ij}}{\lambda_0}\ll 1$. In other words, when the qubits are close enough to one another to experience the bath collectively, or in other words, the bath has a long correlation length (determined by the resonant wavelength $\lambda_0$) relative to the inter-qubit spacing  $r_{ij}$, collective decoherence occurs. $\Gamma_i = \frac{\omega_{i} \mu_{i}^2}{3\pi \epsilon \hbar c^3}$, is the spontaneous emission rate, and
\begin{align}
\Gamma_{ij} &= \Gamma_{ji} = \sqrt{\Gamma_i \Gamma_j} F(k_0 r_{ij}) \text{ for } (i\ne j), 
\end{align}
is the collective incoherent effect due to the dissipative multi-qubit interaction with the bath.
For identical qubits (the case considered here) $\Gamma_{ij}=\Gamma_{ji}$, $\Omega_{ij}=\Omega_{ji}$ and $\Gamma_{i}=\Gamma_{j}$. Solution of Eq.~\eqref{eq:master_equation} for two-qubit initial under the thermal bath is discussed in~\cite{thermalbath,collectivenonclassical}. 

So far, we have discussed the Wigner function, QSL, the phase covariant model, and the two-qubit decoherence model. We now move on to calculate the Wigner function and QSL for the models discussed above. 

\section{Analysis}\label{analysis}
Here, we examine the nonclassicality in the dynamics of a system modeled by phase covariant channel and study the QSL. The speed of evolution is compared with the emergence of nonclassicality. Further, we investigate the presence of nonclassicality in the two-qubit decoherence model under squeezed dynamics both in the independent and collective regimes. Moreover, we study the evolution of quantum correlations using the Werner state as the initial state in the dynamics of the two-qubit decoherence model. 

\subsection{Nonclassicality and quantum speed limit of phase covariant model}
\begin{figure}[h]
    \centering
    \includegraphics[width=1\columnwidth]{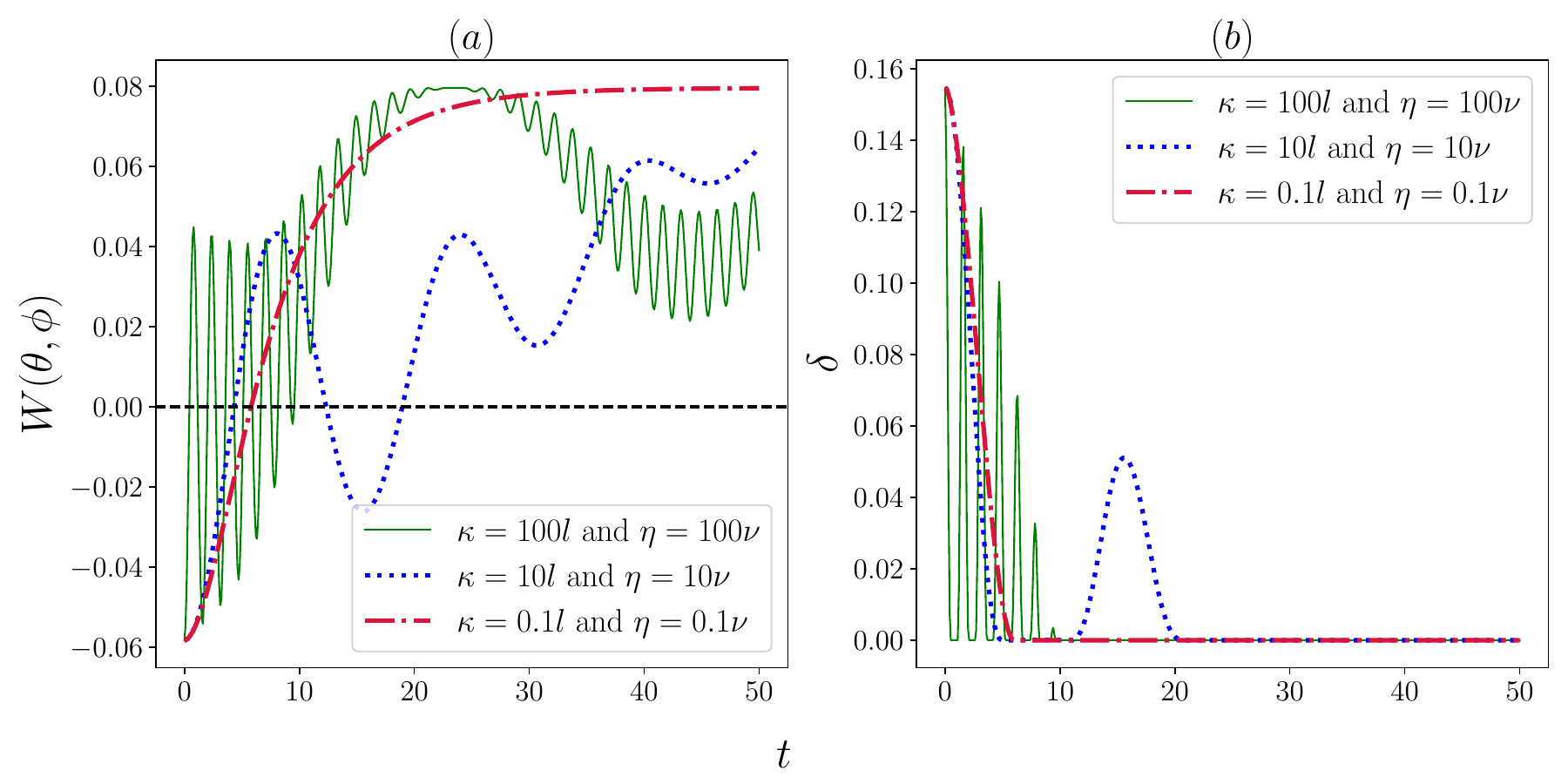}
    \caption{Variation of (a) Wigner function $W(\theta, \phi)$ and (b) nonclassical volume $\delta$ with time for the evolution of a single-qubit system via phase covariant channel. In (a), $\theta = \frac{\pi}3$ and $\phi = \pi$. The initial state is taken to be $\frac{1}2\ket{0} + \frac{\sqrt{3}}{2}\ket{1}$. The solid and dotted lines in both subplots have $l = g = 0.01$, dot-dashed lines in both subplots have $l = g = 1$.}
    \label{fig_W_and_NV_phase_covariant}
\end{figure}

To study the nonclassicality of a system evolving under the phase covariant channel, we use the Wigner function defined in Eq.~(\ref{eq_wigner_function}). In Fig. \ref{fig_W_and_NV_phase_covariant}(a), we plot the Wigner function $W$ with time $t$ for the initial state:  $\frac{1}2\ket{0} + \frac{\sqrt{3}}{2}\ket{1}$, whose evolution is determined by the phase covariant channel, Eq.~(\ref{phase_master_equation}). The density matrix of the single-qubit state at any given time $t$ under the influence of the phase-covariant channel is given by Eq.~(\ref{eq:density}). Specifically, we consider a combination of NMAD and NMRTN channels. The transition from a quantum system to a classical regime can be observed through the behavior of the Wigner function. Initially, the Wigner function of a quantum system exhibits negative values, which are indicative of quantumness. Over time, as the system undergoes decoherence and other processes that lead to classical behavior, the negative regions of the Wigner function decrease. Eventually, the Wigner function becomes entirely positive, signifying the complete transition to classical behavior, where quantum effects are no longer significant.

To quantify the nonclassicality of the system, we use the nonclassical volume given by Eq.~(\ref{nonclassical volume}). Figure~\ref{fig_W_and_NV_phase_covariant}(b) illustrates the temporal evolution of the nonclassical volume $\delta$ for the intiale state $\frac{1}2\ket{0} + \frac{\sqrt{3}}{2}\ket{1}$ under the phase-covariant channel, Eq.~(\ref{phase_master_equation}).
We observe that the nonclassical volume is non-zero during evolution but decreases over time. As shown in the figure, the nonclassical volume goes to zero. However, due to non-Markovian effects, it revives, indicating a temporary return of quantum coherence and interference effects.
Here, we have taken a combination of NMAD and NMRTN channels.
For the NMAD channel, the conditions for non-Markovian behavior is \(l < 2\kappa\), and for the NMRTN channel, the conditions for non-Markovian behavior is \( \left(\frac{2\nu}{\eta}\right)^2> 1 \). The time evolution of the Wigner function, Fig.~\ref{fig_W_and_NV_phase_covariant}(a), clearly demonstrates that in the non-Markovian regime (as indicated by dotted blue and solid green lines), the Wigner function becomes negative multiple times. The frequency of oscillations is very high when $\kappa, \nu \gg l, \eta$. Additionally, the time evolution of nonclassical volume, Fig.~\ref{fig_W_and_NV_phase_covariant}(b), shows that the system's memory effects can revive its nonclassicality. In the Markovian regime (dot-dashed red line), the nonclassical volume decays sharply. 

Next, we study the speed of evolution of the single-qubit system evolving under the phase covariant channel. To this end, we use the QSL $v^{W}_{QSL}$, the form of which is given in Eq.~(\ref{eq_vqsl}). We take the initial state of the single qubit system to be $\frac{1}{\sqrt{2}}\left(\ket{0} + \ket{1}\right)$ and feed this and its time-evolved counterpart's Wigner function into Eq. (\ref{eq_vqsl}) to calculate $v^{W}_{QSL}$. This constitutes the speed of the system's evolution for the given driving time.
\begin{figure}
    \centering
    \includegraphics[width=0.5\textwidth]{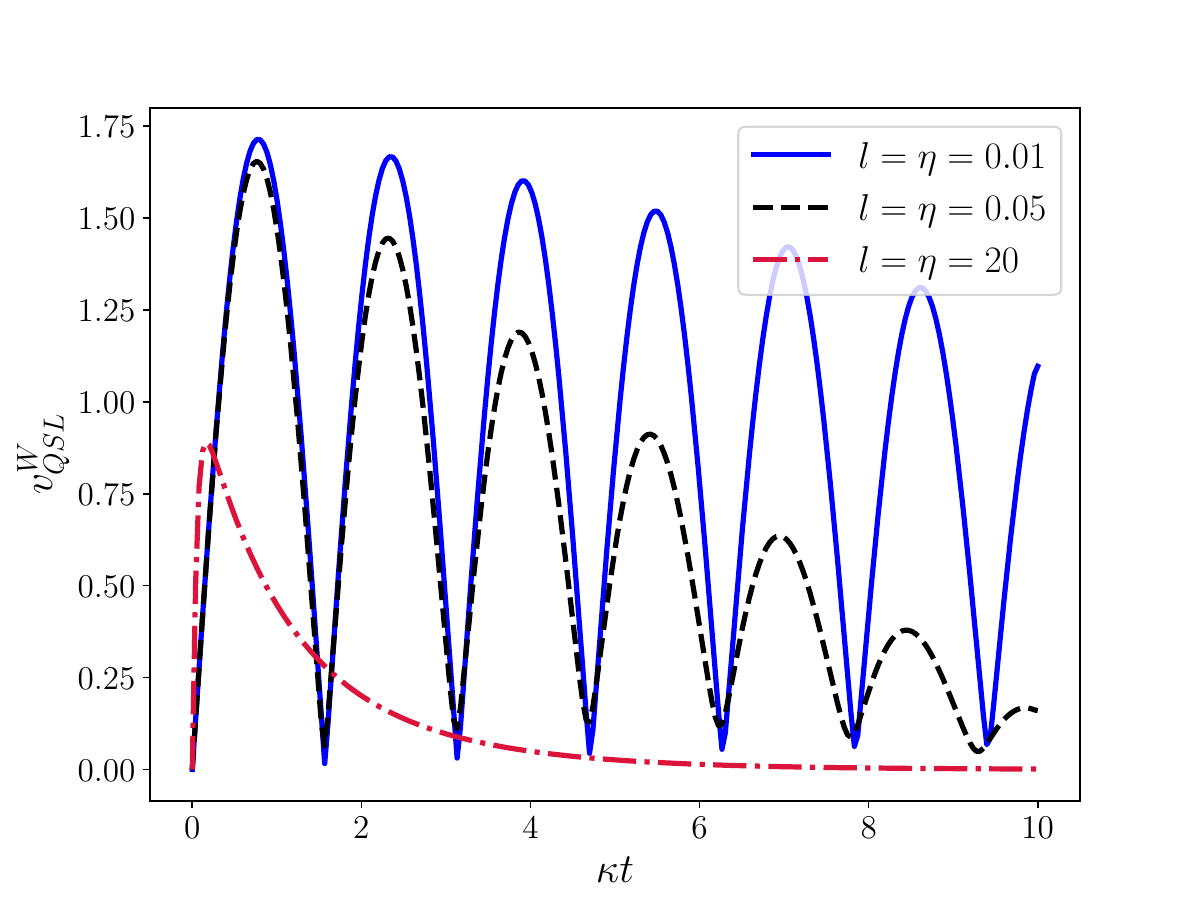}
    \caption{Variation of QSL $v^{W}_{QSL}$ with time for the system's evolution under phase-covariant channel in the (non-)Markovian regime. Here, $\nu = \kappa$ is taken to be one. Further, the initial state is taken to be $\frac{1}{\sqrt{2}}\left(\ket{0} + \ket{1}\right)$.}
    \label{fig_tau_qsl_phase_covariant}
\end{figure}
The variation of $v^{W}_{QSL}$ with time for the evolution of the single-qubit system under phase covariant channel is plotted in Fig.~\ref{fig_tau_qsl_phase_covariant}. In the non-Markovian regimes (depicted by solid blue and dashed black lines), when $2\kappa, 2\nu \gg l, \eta$, the system's speed of evolution is very fast and oscillatory in nature.
Upon increasing the values of $\eta$ and $l$, the speed of evolution slows down as $v^{W}_{QSL}$ gets lower values, and the local maxima values also reduce as time progresses. Further, in the Markovian regime (depicted by the dot-dashed red line), the system's speed of evolution decreases rapidly as $v^{W}_{QSL}$ goes to zero with time. This brings out that non-Markovianity aids in speeding up the evolution.
\begin{figure}
    \centering
    \includegraphics[width=1\columnwidth]{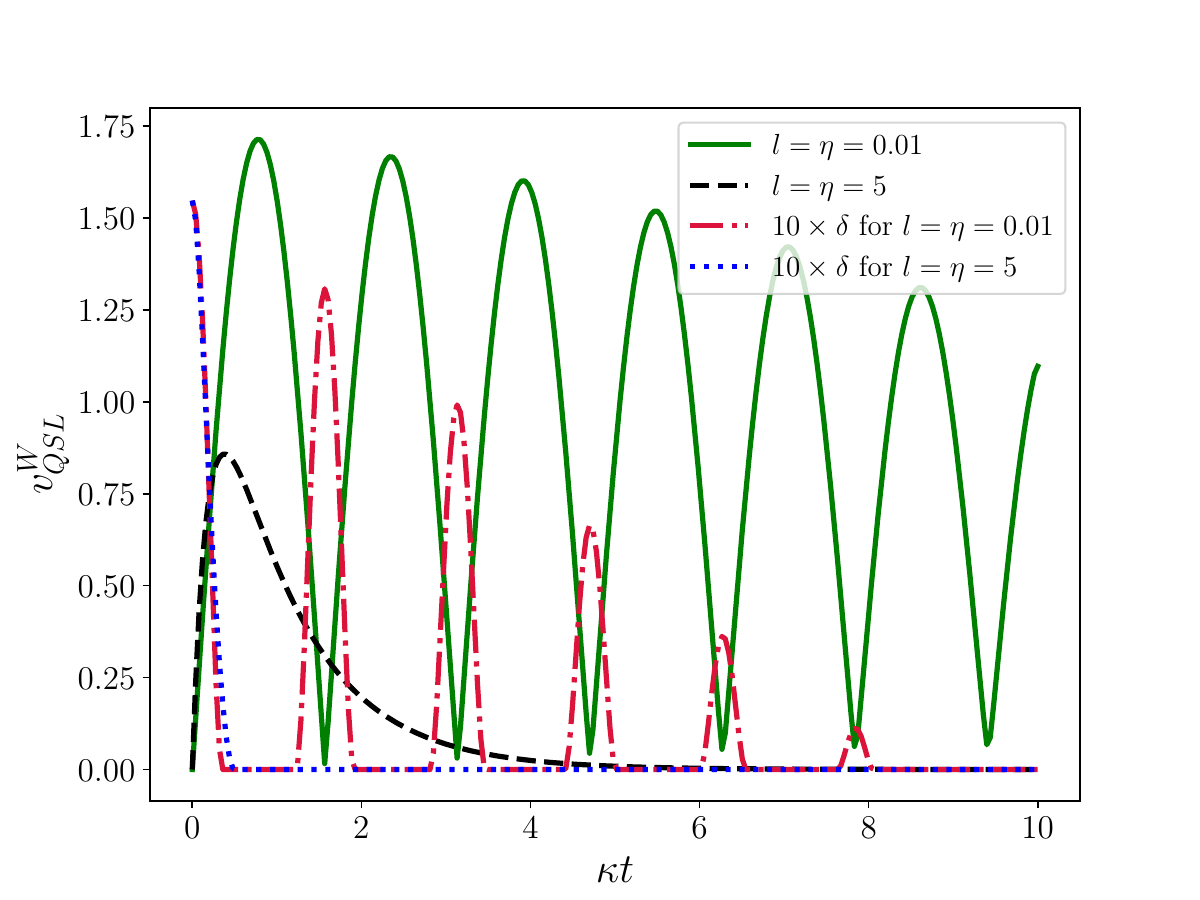}
    \caption{Variation of the QSL $v^{W}_{QSL}$ and the scaled nonclassical volume $10\times\delta$ with time for the system's evolution under the phase-covariant channel. Here, $\nu = \kappa$ is taken to be one, and the initial state is taken to be $\frac{1}{\sqrt{2}}\left(\ket{0} + \ket{1}\right)$.}
    \label{fig_delta_qsl_phase_covariant}
\end{figure}

Further, to compare the nonclassicality in the system with the speed of evolution, we plot the nonclassical volume and the QSL together in Fig.~\ref{fig_delta_qsl_phase_covariant}. It can be observed that the peaks of the nonclassical volume match with the valleys of the QSL. This brings out a relationship between the nonclassical volume and the QSL. The rise in the speed of evolution can be seen as the effect of non-zero nonclassical volume; when it becomes zero, the speed of evolution starts decreasing. Further, in the Markovian regime, no revival can be seen in both the speed of evolution and non-classical volume.

\subsection{QSL for the evolution of the two-qubit dissipative model using Wigner function}
Here, we analyze the Wigner function, non-classical volume, and the evolution of the QSL for a two-qubit initial separable state evolving under a squeezed thermal bath, as discussed above. We take the squeezing angle to be $\phi=0$, and $k_0=1$, $\omega_0=1$, $\Gamma_j=0.05$ and $\mu.r_{ij}=0$.

\begin{figure}
    \centering
    \includegraphics[width=0.5\textwidth]{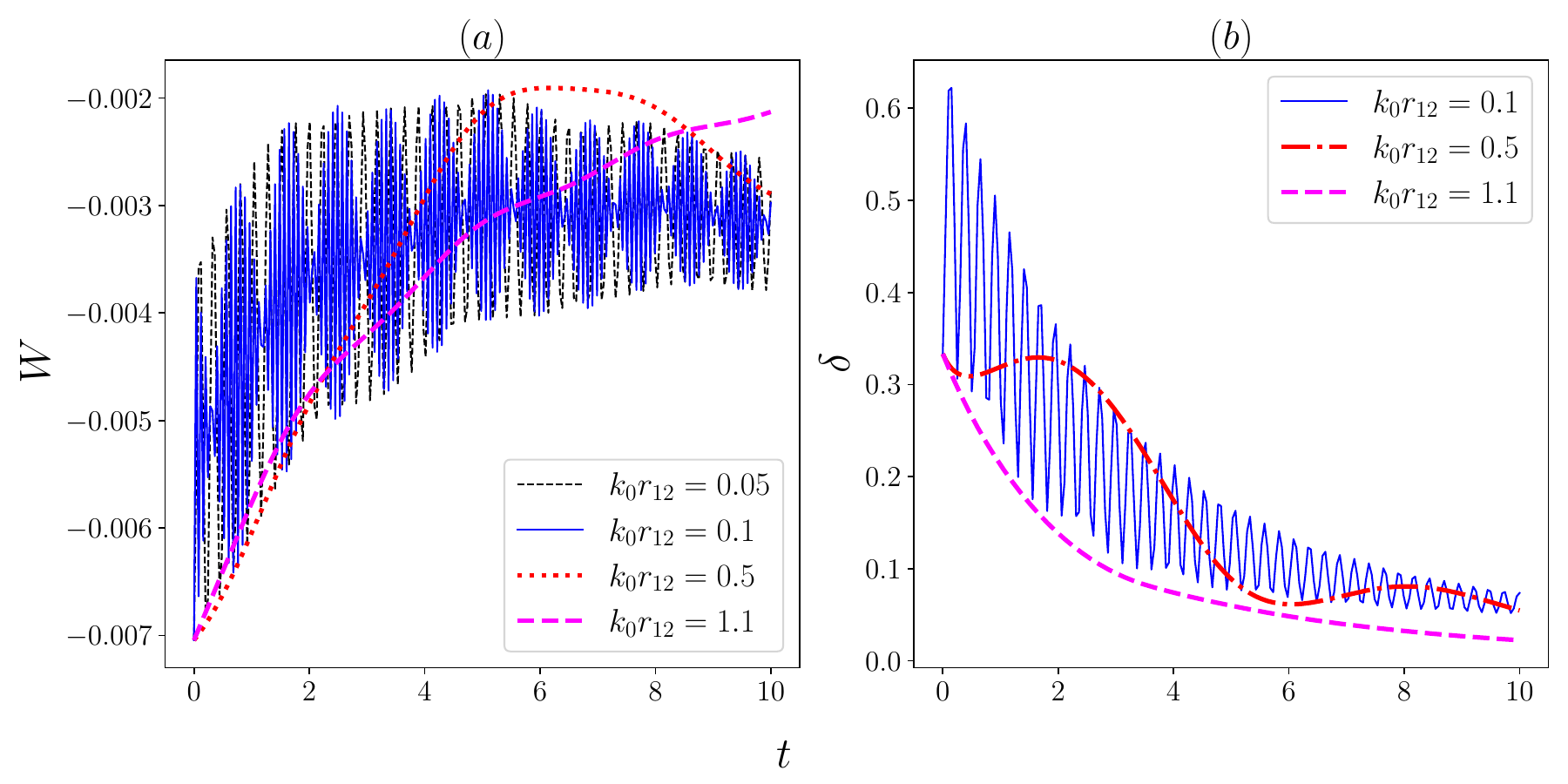}
    \caption{Variation of the (a) Wigner function $W(t)$ and (b) nonclassical volume $\delta$ with time $t$ for a two-qubit state ($\ket{01}$) under a squeezed thermal bath. Here, we have taken $\theta_1=\frac{\pi}{4}$, $\theta_2=\frac{\pi}{6}$, $\phi_1=\frac{\pi}{8}$, $\phi_2=\frac{\pi}{6}$ for (a). For both (a) and (b), temperature $T=1$, bath squeezing parameters $r=-0.2$ and $\phi=0$. Further, $\omega_1 = \omega_2 = 1$ and $\Gamma_1 = \Gamma_2 = 0.05$.}
    \label{fig_W_and_NV_2_qubit_model}
\end{figure}

Figure~\ref{fig_W_and_NV_2_qubit_model}(a) illustrates the time-dependent variation of the Wigner function for a two-qubit initial state  $\ket{01}$ under the influence of a squeezed thermal bath at different inter-qubit distances. We observe that the Wigner function exhibits negative and oscillatory behavior at small inter-qubit distances (collective regime). However, as the inter-qubit distance increases, the Wigner function oscillations cease, though they still remain negative. Figure~\ref{fig_W_and_NV_2_qubit_model}(b) depicts the variation of the nonclassical volume $\delta$ as a function of time for the evolution of a two-qubit initial separable state mentioned above. The plot reveals that the nonclassical volume oscillates and decreases over time. However, as the inter-qubit distance increases, these oscillations diminish. This behavior is similar to the behavior of the Wigner function under similar conditions. 

\begin{figure}
    \centering
    \includegraphics[width=0.5\textwidth]{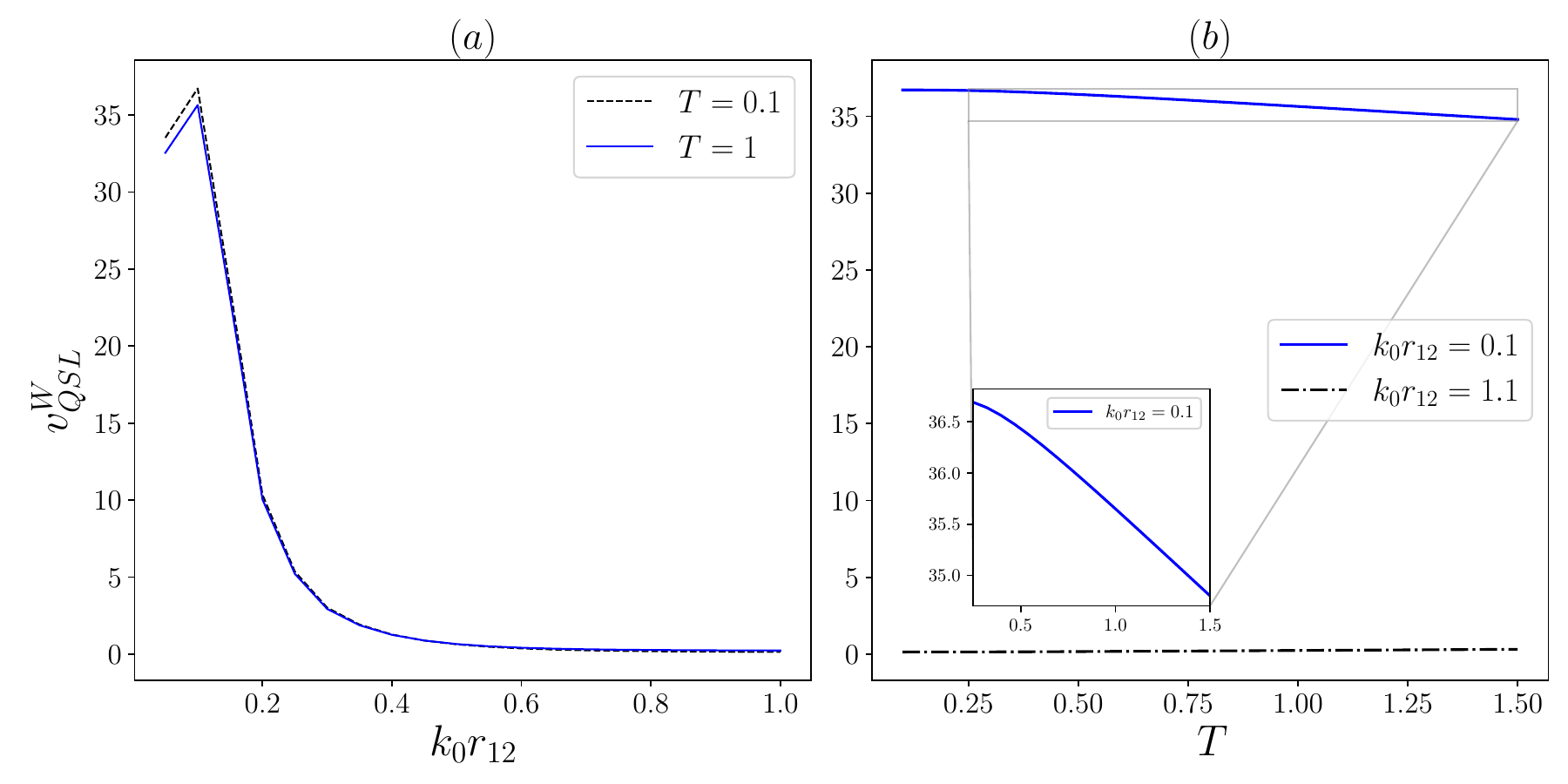}
    \caption{Variation of QSL $v^{W}_{QSL}$ with (a) $k_0r_{12}$ at two different temperatures $T=0.1$ and $T=1$ and (b) $T$ for two inter-qubit distances $k_0r_{12} = 0.1$ and $k_0r_{12} = 1.1$. The initial two-qubit state ($\ket{01}$) is evolved under a squeezed thermal bath with bath squeezing parameters $r=-0.2$ and $\phi=0$. The time is taken to be $t=0.5$. The other parameters are taken to be $\omega_1 = \omega_2 = 1$ and $\Gamma_1 = \Gamma_2 = 0.05$. In (b), the black dot-dashed curve ranges from 0.14 to 0.3, which appears to be around zero in comparison to the blue plot, which has values close to 35.}
    \label{fig_qsl_2_qubit_model}
\end{figure}

In Fig.~\ref{fig_qsl_2_qubit_model}(a), we plot the variation of the QSL as a function of the inter-qubit distance for an initial state $\ket{01}$ under a squeezed thermal bath at different temperatures. The QSL shows a similar pattern of peaks and valleys for different temperatures. However, the values of QSL are slightly higher for the lower temperature. The corresponding variation as a function of temperature for different inter-qubit distances is plotted in Fig.~\ref{fig_qsl_2_qubit_model}(b). As evident, the evolution is faster in the collective regime (bold curve) as compared to that in the independent regime (dot-dashed curve). Further, the speed of evolution slows down as we increase the temperature. It is known that entanglement is generated from the separable state in this model in the collective regime~\cite{thermalbath, Banerjee2009}. The above plots benchmark the fact that due to the generation of the entanglement, the speed of the evolution is faster~\cite{Devvrat_QSL}.

To study the impact of the evolution on the correlation between the two qubits, we study the quantum discord together with the QSL. To this end, we take the Werner state as the system's initial state. The two-qubit Werner state is given by~\cite{Werner}
\begin{align}
   \rho_{W_{AB}} = P|\psi\rangle\langle\psi|_{AB}+\frac{1-P}{4}\mathbb{I}_{AB},
\end{align}
where $|\psi\rangle_{AB}=\frac{1}{\sqrt{2}}\left(|0\rangle_A|0\rangle_B + |1\rangle_A|1\rangle_B\right)$ and $P$ is the mixing probability such that $0\leq P \leq 1$. Furthermore, we study the quantum correlations using quantum discord~\cite{Zurek_discord}, which is defined as 
\begin{align}
\mathit{D}\left(A:B\right)=S\left(B\right)-S\left(A,B\right)+S\left(A|B\right),
\end{align}
where $S(A)$ and $S(B)$ are the von Neumann entropies of the subsystem states $\rho_A$ and $\rho_B$, respectively. The terms $S(A, B)$ and $S(A|B)$ are the joint von Neumann entropy and the quantum conditional entropy of the system, respectively.
The quantum conditional entropy $S(A|B)$ is given by
\begin{align}
    S\left(A|B\right) =\min_{\{\Pi_k\}}  \sum_{k=1}^{\mathit{H}_B} p_k S(\rho_{A|\Pi_k} ),
\end{align}
where $\mathit{H}_B$ is the Hilbert space dimension of subsystem $B$. $\rho_{A|\Pi_k}$ is the post-measurement state for the subsystem $A$ when a measurement is performed on the subsystem $B$ and $p_k=\text{Tr}({\Pi_k}^{\dagger}\Pi_k\rho_{AB})$ is the probability associated with the measurement operators $\Pi_k$. $\rho_{A|\Pi_k}$ can be explicitly written as
\begin{align}
    \rho_{A|\Pi_k} = \frac{1}{p_k}\text{Tr}_B\left(\Pi_k\rho_{AB}\Pi_k\right).
\end{align}

\begin{figure}
    \centering
    \includegraphics[width=0.5\textwidth]{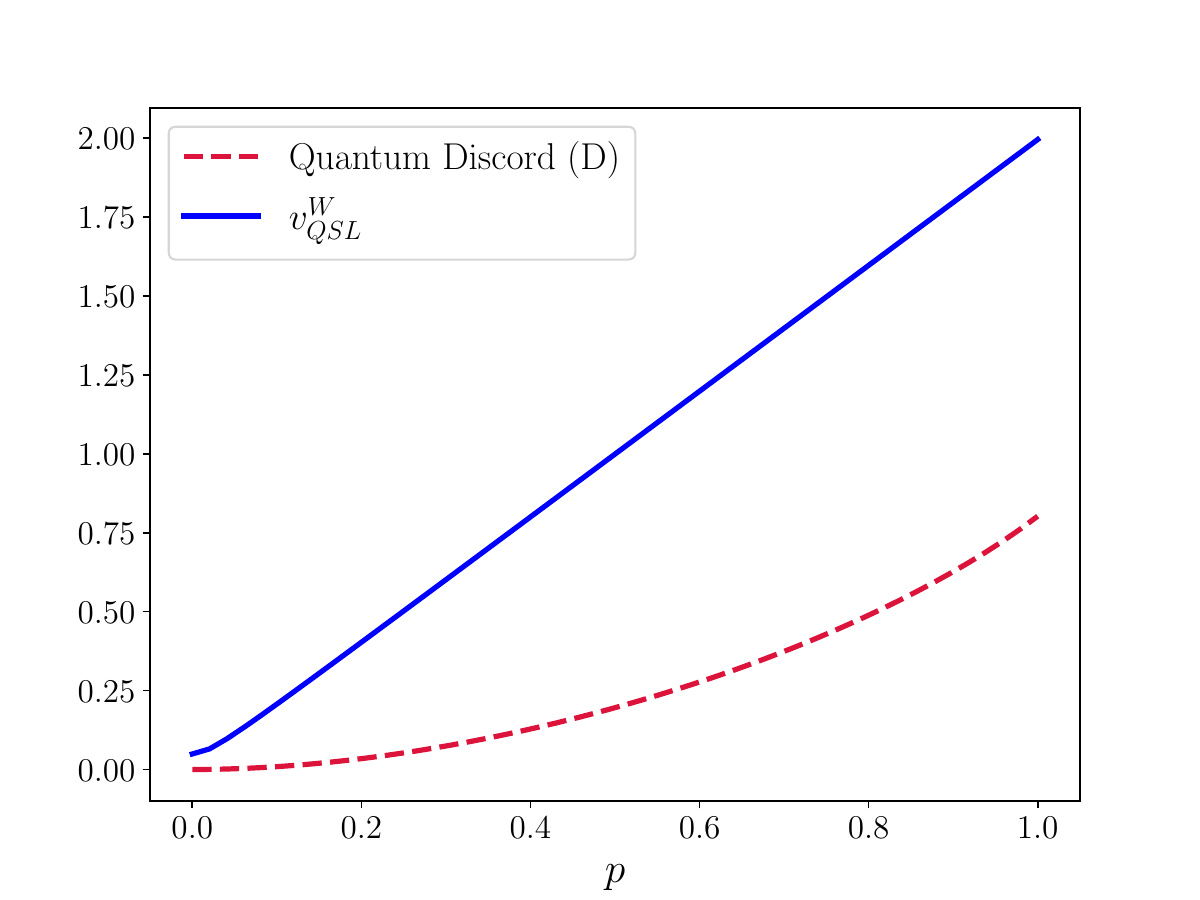}
    \caption{Variation of QSL $v^{W}_{QSL}$ and quantum discord $D$ with mixing parameter $P$ for the Werner state under a squeezed thermal bath with bath squeezing parameters $r=0.5$ and $\phi=0$ for inter-atomic distances $k_0r_{12}=0.05$, $T=1$ and $t=1$. }
    \label{fig:sample8}
\end{figure}

We calculate the quantum discord and QSL as a function of the mixing probability $P$ and plot them in Fig.~\ref{fig:sample8}. The QSL is observed to be increasing with the value of the $P$. It has the lowest value when the state is a separable maximally mixed state and the highest value when it is maximally entangled ($P=1$). This reiterates that the speed of evolution is faster when quantum correlations increase, which is consistent with what was observed in~\cite{Devvrat_QSL}. Quantum discord shows a monotonically increasing behavior. At $P = 1$, the Werner state becomes the Bell state, which should have the maximum value of discord, that is, one. However, due to the interaction of the thermal bath with the two-qubit system, the value of quantum discord is lower than one at $P=1$. 

\section{Conclusions}\label{conclusion}
Here, we focused on the quantum speed limit using the Wigner function. This had the advantage of computational simplicity and provided a visualization of quantum to classical transition. A single-qubit system evolving under the phase covariant channel and a two-qubit system interacting with a squeezed thermal bath with position-dependent coupling were examined. We investigated the Wigner function, non-classical volume, and speed limit for both models. The negative values of the Wigner function and non-zero nonclassical volume depicted the quantumness of the above systems. 
The non-Markovian region in the phase covariant model was found to favor the rebirth of nonclassicality in the system as well as the system's speed of evolution. The nonclassical volume was seen to be related to the system's speed of evolution.
The Wigner function and the nonclassical volume were observed to be oscillatory in the collective regime of the two-qubit model.  
The inter-qubit distance in the two-qubit dissipative model had an interesting role in the system's evolution. It was observed for the two-qubit model that the collective regime was conducive to speeding up the evolution. This could be attributed to the generation of quantum correlations in this regime, such as entanglement and quantum discord. Further, with an increment in the temperature, the system slowed down. 

\bibliographystyle{apsrev4-1}
\bibliography{reference}  

\end{document}